\documentclass[pra,twocolumn]{revtex4-1}
\usepackage{hyperref}
\usepackage{graphicx}
\begin{document}

\title{Corrections to the expected signal in quantum metrology using highly anisotropic Bose-Einstein Condensates}
\author{Salini Jose}
\affiliation{School of Physics, Indian Institute of Science Education and Research Thiruvananthapuram, Kerala, India 695016}
\email{salini@iisertvm.ac.in}
\author{Anil Shaji}
\affiliation{School of Physics, Indian Institute of Science Education and Research Thiruvananthapuram, Kerala, India 695016}
\email{shaji@iisertvm.ac.in}

\begin{abstract}
In the quantum metrology protocol described by Tacla et al.~[Tacla et~al., Phys.~Rev.~A {\bf 82}, 053636 (2010)] where a two mode Bose-Einstein condensate (BEC) is used for parameter estimation, the measured quantity is to be obtained by doing a one parameter fit of the observed data to a theoretically expected signal. Here we look at different levels of approximation used to model the two mode BEC to see how the estimate improves when increasing level of detail is added to the theory while at the same time keeping the expected signal computable.
\end{abstract}
\maketitle
 
\section{Introduction \label{sec1}}

Nonlinear interactions between the $N$ elementary units of the quantum probe in particle based quantum limited single parameter estimation schemes can allow the measurement uncertainty to scale as $1/N^{k}$, where $k$ is the degree of nonlinearity~\cite{boixo_generalized_2007}. Without loss of generality the elementary units that make up the quantum probe can be taken to be qubits. The scaling of the minimum possible uncertainty in the value of the classical parameter that is measured with respect to the resources that go into the metrology protocol, as quantified by $N$, is a measure of the efficiency and performance of the measurement scheme. The ``Heisenberg limited scaling'' of $1/N$ was held to be the absolute limit for the performance of a particle based metrology scheme like Ramsey interferometry \cite{bollinger,huelga} until it was clarified that this is so under the assumption that each of the probe qubits undergoes independent parameter dependent evolutions. 

With nonlinear parameter dependent couplings between the probe qubits it was shown that even if the initial state of the quantum probe is restricted to being a product state of the $N$ units, the best possible scaling of the uncertainty is $1/N^{k-1/2}$~\cite{boixo_quantum-limited_2007}.  In a sequence of related publications~\cite{boixo_quantum_2008,boixo_quantum-limited_2009,Boixo-Tacla} a proposal to use a two mode BEC to perform a proof of principle experiment that demonstrates a scaling of $1/N^{3/2}$ was put forward. Realizations of the experiment in the lab are also being successfully pursued~\cite{egorov_long-lived_2011,egorov_measurement_2013,napolitano_interaction-based_2011}. The effective nonlinear interaction modeled as the $g|\psi|^{2}$ potential term in the Gross-Pitaevski equation~\cite{Dalfovo,legget} describing the condensate under the mean field approximation furnishes a $k=2$ coupling that can be used to design an experiment that estimates the value of a function of the  constant $g$ with the measurement uncertainty potentially scaling as $1/N^{3/2}$. 

The initial proposal in~\cite{boixo_quantum_2008} was developed further in~\cite{boixo_quantum-limited_2009},~\cite{Boixo-Tacla} as well as in~\cite{taclacaves} to include incrementally the deleterious effects of possible non-ideal conditions in the lab as well as to relax the simplifying assumptions that were made in the theoretical analysis. This Paper is also another step in the same direction where we focus on one of the simplifying assumptions made about the initial state of the BEC and explore how and when corrections due to the relaxation of this assumption that were proposed in~\cite{taclacaves} would come into play. 

The motivation for refining the original proposal of the BEC based metrology scheme is to push towards a viable experiment in the lab. In a nutshell the problem can be posed as follows: under non-ideal conditions with none of the simplifying assumptions the effective scaling we get is $1/(N^{3/2}\eta_{N})$ where $\eta_{N}$ can depend on $N$ as well as have higher order dependencies on the measured parameter itself. Enumerating and understanding all the factors that influence $\eta_{N}$ and crucially its $N$ dependence is important for a fool proof interpretation of the proposed experiment as one that surpasses the $1/N$ scaling. The same problem can be viewed from a slightly different perspective also. The signal that is measured in the final step of the experiment is the number of atoms in each one of the two internal states of the two mode BEC. Repeating the experiment a fixed number of times gives us the time dependence of these numbers. In the ideal situation the only unknown quantity on which the time dependence of the population of atoms depends on is the measured parameter. However, in practice it would depend on $\eta_{N}$ as well. Therefore if the strategy of doing a one parameter fit of the measured time evolution of the populations so as to find the value of the parameter is to work, an almost complete knowledge of the dependence of $\eta_{N}$ on the details of the experiment is essential. 

\section{Quantum limited measurements using a two mode BEC \label{sec2}}

 In~\cite{boixo_quantum_2008} and~\cite{boixo_quantum-limited_2009}, a two mode BEC at zero temperature is the quantum probe. Limiting the possible electronic (hyperfine) states of the atoms to just two lets us treat each one as a  qubit. At zero temperature all the atoms can be assumed to be in the ground state of the condensate with identical overlapping wave functions. The mean field approximation holds under these conditions and if we further assume that all the atoms are initially in one of the two possible hyperfine states then their wave function is given by the ground state solution of the time-inedependent Gross Pitaevski ~\cite{Dalfovo,legget} equation,
 \begin{equation}
 	\label{eq:cgp1}
	  \bigg[ - \frac{\hbar^{2}}{2m} \nabla^{2} + V + g (N-1)  |\psi_{N}|^{2} \bigg] \psi_{N} = \mu_{N} \psi_{N},
 \end{equation}
where $N$ is the number of atoms in the BEC, $\mu_{N}$ is the chemical potential, $V$ is the external trapping potential and the coupling constant $g$ is related to the $s$-wave scattering length $a$ and the mass $m$ of the atoms as 
\begin{equation}
	\label{eq:const1}
	g=\frac{ 4 \pi \hbar^{2} a}{m}.
\end{equation}  

 The proposed experiment proceeds by applying an instantaneous pulse that puts each atom in the condensate in a specific superposition of the two internal states. Assuming that only elastic collisions are allowed between the atoms in the two hyperfine states labeled by $|1\rangle$ and $|2\rangle$ respectively,  the Hamiltonian of the system is, 
 \begin{eqnarray}
 	\label{eq:hamil1}
	\hat{H} & = & \sum_{\alpha =1,2} \int d {\mathbf r} \, \bigg[ \frac{\hbar^{2}}{2m} \nabla \hat{\psi}^{\dagger}_{\alpha} \cdot \nabla \hat{\psi}_{\alpha}^{\vphantom{\dagger}} + V(r)\hat{\psi}^{\dagger}_{\alpha} \hat{\psi}_{\alpha} ^{\vphantom{\dagger}} \bigg] \nonumber \\
	&& \qquad + \; \frac{1}{2} \sum_{\alpha, \beta} g_{\alpha \beta} \int d{\mathbf r} \, \hat{\psi}^{\dagger}_{\beta} \hat{\psi}^{\dagger}_{\alpha}  \hat{\psi}_{\alpha}^{\vphantom{\dagger}}\hat{\psi}_{\beta}^{\vphantom{\dagger}},
 \end{eqnarray}
 where $\hat{\psi}_{\alpha}$ is the modal annihilation operator. For a zero temperature BEC we can truncate the modal annihilation operator to just one term of the form
 \[ \hat{\psi}_{\alpha} = \psi_{N, \alpha} ({\mathbf r}) \hat{a}_{\alpha}. \]

In~\cite{boixo_quantum_2008} the rather strong assumption that at least for short times after the pulse that puts the atoms in a superposition state, the spatial part of their wave functions are identical, is made so that
\[ \hat{\psi}_{\alpha} = \psi_{N} ({\mathbf r}) \hat{a}_{\alpha}. \]
With this assumption it was shown that the Hamiltonian in Eq.~(\ref{eq:hamil1}) can be brought to the form
\begin{equation}
	\label{eq:hamil2}
	\hat{H} = H_{0} + \gamma_{1} \eta_{N} (N-1) \hat{J}_{z} + \gamma_{2} \eta_{N} \hat{J}_{z}^{2},
\end{equation}
where
\[ \hat{J}_{z} = \frac{1}{2} (\hat{a}_{1}^{\dagger} \hat{a}_{1}^{\vphantom{\dagger}} - \hat{a}_{2}^{\dagger} \hat{a}_{2}^{\vphantom{\dagger}} ),  \]
and
\begin{equation} 
	\label{eq:eta1}
	\eta_{N} = \int d{\mathbf r} |\psi_{N}({\mathbf r})|^{4}. 
\end{equation}
$H_{0}$ is a $c$-number energy that depends on $N$ while the constants $\gamma_{1}$ and $\gamma_{2}$ characterize the three elastic scattering processes listed above and are defined as
\[ \gamma_{1} \equiv \frac{1}{2} ( g_{11} - g_{22}) \quad {\rm and} \quad \gamma_{2} \equiv \frac{1}{2} (g_{11} + g_{22}) - g_{12}, \]
with $g_{ij}$ being related to the scattering length $a_{ij}$ of the corresponding process through Eq.~(\ref{eq:const1}). 

The proposed experiment uses $^{87}\,{\rm Rb}$ atoms in the $|F=1; M_{F} =-1\rangle \equiv |1\rangle$ and $|F=2; M_{F} =+1\rangle \equiv |2\rangle$ states. This choice has the added advantage that the ratios $\{a_{22}:a_{12}:a_{11} \} = \{ 0.97:1:1.03\}$ ~\cite{williams} mean that 
\[ \gamma_{2} = 0. \]
So for a two mode BEC of $^{87}\,{\rm Rb}$ atoms, $\gamma_{1} \eta_{N} (N-1) \hat{J}_{z} $ is the only term in the Hamiltonian in Eq.~(\ref{eq:hamil2}) that generates a relative phase between the two states $|1\rangle$ and $|2\rangle$. This phase can be used to estimate the value of the parameter $\gamma_{1}$ from which we can obtain the value of one of the two scattering lengths, say $a_{22}$, given that we know $a_{11}$. Since $\gamma_{1}$ and $a_{22}$ are linearly related through $g_{22}$ with the scaling of the measurement uncertainty in both being identical, in the following, we will consider $\gamma_{1}$ as the measured parameter but we will keep in mind that $a_{11}$ (and $g_{11}$) are known quantities. Comparing with conventional Ramsey interferometry \cite{Gleyzes} in which the parameter dependent evolution of the probe is generated by a Hamiltonian proportional to $J_{z}$ we can see how the relative phase evolves $N$ times faster (assuming $N\gg 1$). As detailed in~\cite{boixo_quantum-limited_2009} the quantum Cramer-Rao bound on the measurement uncertainty in $\gamma_{1}$ will scale with $N$ as
\[ \delta \gamma_{1} \sim \frac{1}{\eta_{N} N^{3/2}}. \]
 Achieving a $1/N^{3/2}$ scaling requires $\eta_{N}$ not to depend on $N$. From the definition of $\eta_{N}$ in Eq.~(\ref{eq:eta1}) we see that $\eta_{N}^{-1}$ is proportional to the volume of the ground-state wave function.  The primary reason for the dependence of $\eta_{N}$ on $N$ is that as the number of atoms increases the BEC ground state wave function expands. To avoid the expansion of BEC wave function, in~\cite{boixo_quantum_2008} highly anisotropic traps are proposed so that the expansion will be along only a few of the dimensions at least over a range  of $N$ that is sufficient to demonstrate a scaling better than $1/N$. It is known that three-dimensional Bose-Einstein condensates confined in highly anisotropic traps exhibit lower dimensional behavior when the number of condensed atoms is well below a critical value~\cite{gorlitz}. Under such conditions, the expansion of the BEC wave function with $N$ along the tightly confined dimensions can be effectively neglected because the characteristic energy scale along those directions far exceeds the scattering energy of the atomic cloud.
 
Repeating the initial pulse that put the $^{87}\,{\rm Rb}$ atoms in an equal superposition of the two hyperfine levels would convert the relative phase information between the two states into population information. If the populations of atoms in states $|1\rangle$ and $|2 \rangle$ are measured as a function of time  through multiple repetitions of the experiment, we expect it to oscillate in time with a frequency $\Omega_{N} = (N-1) \eta_{N} \gamma_{1}/\hbar$ provided all the assumptions detailed above hold.  If $\eta_{N}$ is known along with $N$ then a one parameter fit of the observed population versus time data will give us the value of the measured parameter $\gamma_{1}$. 

As mentioned before, to model the expected signal from a real experiment, we have to relax the assumptions made previously and obtain a more detailed expression for the time dependence of the two populations. In the next section we discuss the approaches for obtaining an expression for the time dependence.

\section{Time dependence of the atomic populations and estimating \texorpdfstring{$\eta_{N}$}{eta} \label{Sec3}}

Under the mean field approximation the dynamics of the two mode BEC initialized in the state $(|1\rangle + |2\rangle)/\sqrt{2}$ is governed by the time-dependent, coupled GP equations, 
\begin{equation}
	\label{eq:gp2}
	i \hbar \frac{\partial \psi_{N, \alpha}}{\partial t} = \bigg(\!\! -\frac{\hbar^{2}}{2m} \nabla^{2} + V + \frac{N-1}{2} \sum_{\beta} g_{\alpha \beta} |\psi_{N, \beta}|^{2} \!\!\bigg) \psi_{N, \alpha},
\end{equation}
with $\alpha, \beta = 1,2$. In the absence of actual experimental data, we will compare the theoretically expected signal to the atomic populations obtained by numerical integration of the above pair of equations. It must be noted that for this reason, we will not be going beyond the mean field approximation in this Paper and also that we put in the known value of $a_{22}$ into the numerical integration of the coupled GP equation and in this sense, the data obtained is equivalent to the data from a true experiment. 

The population of atoms in each of the two hyperfine states is given by~\cite{boixo_quantum-limited_2009}
\begin{equation}
	\label{eq:pop}
	p_{1,2} = \frac{1}{2} [ 1 \pm {\rm Im} (\langle \psi_{N,1} | \psi_{N,2} \rangle)],
\end{equation}
and is therefore determined by the overlap of the two spatial wave functions, 
\begin{equation}
	\label{eq:Ot}
	O(t) = \langle \psi_{N,1} | \psi_{N,2} \rangle = \int d^{3} r \, \psi_{N,1}^{*}({\mathbf r}, t) \psi_{N,2}({\mathbf r}, t) .
\end{equation} 
This overlap will be the main quantity of interest in the rest of the paper. In the simplest and rather ideal case where the spatial part of the two wave functions are assumed to be identical except for a spatially independent relative phase between the two, we have 
\[ p_{1,2} = \frac{1}{2}(1 \pm \sin \Omega_{N}t), \qquad \Omega_{N} = (N-1) \eta_{N} \gamma_{1}/\hbar, \]
as discussed previously.

In order to estimate $\eta_{N}$, in~\cite{boixo_quantum-limited_2009} for an anisotropic trap a product form for the ground state wave function of the condensed atoms is assumed as
\begin{equation}
	\label{eq:gnd1}
	\psi_{0}(\rho, \, z) = \chi_{0}(\rho) \phi_{N}(z),
\end{equation}
where $z$ labels the $d$ loosely confined ``longitudinal'' dimensions while $\rho$ labels the remaining $D=3-d$ tightly confined ``transverse'' dimensions. The main objective of using an anisotropic trap is to find a working range of $N$ for which the spreading of the wave function as $N$ increases is significant only along the longitudinal dimensions thereby limiting the change in $\eta_{N}$ with $N$ and yielding a better than $1/N$ scaling for $\delta \gamma_{1}$. In~\cite{boixo_quantum-limited_2009} two critical atom numbers $N_{L}$ and $N_{T}$ that bound this working range is identified. 

For concreteness, from here on, we will restrict our discussion to a quasi-one dimensional BEC in harmonic trapping potentials so that $d=1$. If $N$ is in the effective working range between $N_{L}$ and $N_{T}$ we may assume that the transverse part of the product wave function in Eq.~(\ref{eq:gnd1}) is just the ground state wave function of the transverse harmonic trapping potential,  
\begin{equation}
	\label{eq:trans1}
	\chi_0=\frac{1}{\sqrt{2 \pi\rho_0^2}}\exp \bigg(-\frac{\rho^2}{4 \rho_0^2} \bigg),
\end{equation}
where $\rho_{0} = \sqrt{\hbar/2 m \omega_{T}}$ with $\omega_{T}$ being the frequency of the transverse trap.  The product wave function assumed in Eq.~(\ref{eq:gnd1}) lets us split $\eta_{N}$ also into a product as
\begin{equation}
	\label{eq:eta1a}
	\eta_{N} = \eta_{T} \eta_{L}.
\end{equation}
Using $\chi_{0}$ from Eq.~(\ref{eq:trans1}) above, we get 
\begin{equation}
	\label{eq:etat1}
	\eta_{T} =\frac{1}{4 \pi \rho_{0}^{2}}. 
\end{equation}
We can now use $\eta_{T}$ in the one dimensional reduced, time-independent GP equation, 
\begin{equation}
	\label{eq:redGP1}
	\bigg[-\frac{\hbar^2}{2m}\frac{d^2}{dz^2}+\frac{1}{2}m \omega_{L}^{2} z^2+g_{11}(N-1)\eta_T|\phi_{N}|^2 \bigg]\phi_{N}=\mu_L\phi_{N},
\end{equation}
to obtain the longitudinal wave function $\phi_{N}(z)$. Here $\mu_{L} = \mu_{N} - \hbar \omega_{T}$ is the longitudinal part of the  chemical potential. We can compute $\eta_{L}$ and its $N$ dependence from $\phi_{N}$.

\begin{figure}[!htb]
\resizebox{8 cm}{5 cm}{\includegraphics{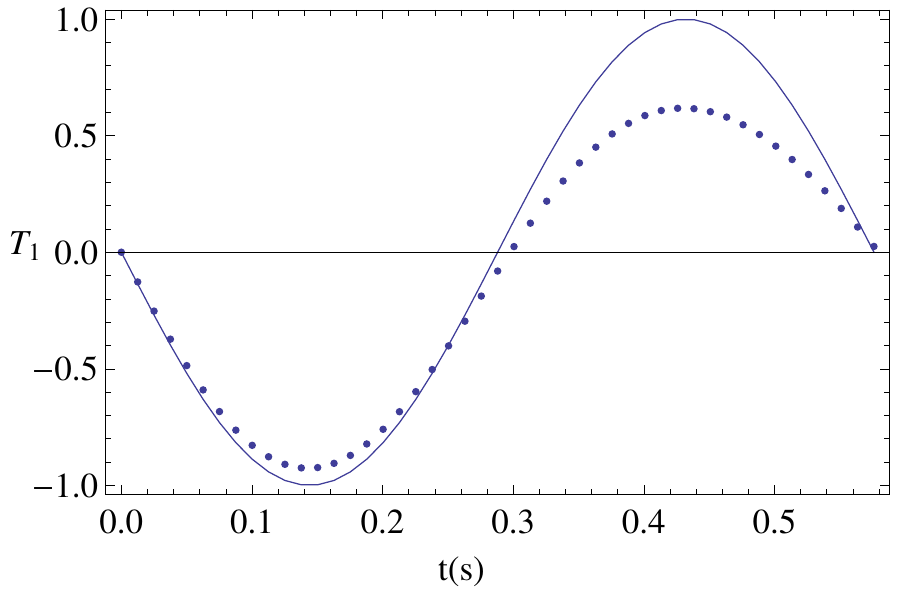}}
\resizebox{8 cm}{5 cm}{\includegraphics{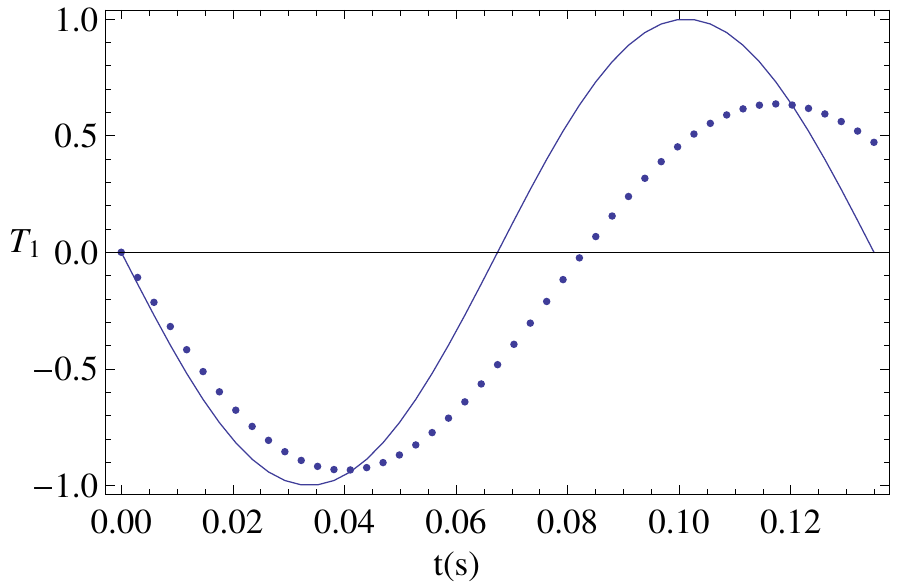}}
\caption{(color online) The numerically computed values of $O_{I}(t)$ are shown with dotted lines while $T_{1}(t)$ is the solid line in each of the two graphs. The figure on top is for $N=1000$ and the one below for $N=10,000$. The trapping frequencies used are 350Hz along the transverse dimensions and 3.5Hz along the longitudinal dimensions \label{fig:figure1}}
\end{figure}

When $N$ is much larger than $N_L$ but still smaller than $N_{T}$, the kinetic-energy term in the reduced GP equation can be neglected, and the longitudinal wave function is approximated by the Thomas-Fermi solution \cite{legget}
\begin{equation}
\label{eq:lwave}
|\phi_N(z)|^2=\frac{\mu_L-m \omega_{L}^{2} z^2/2}{(N-1)g_{11}\eta_T}
\end{equation}
Using normalization condition for $\phi_{N}$  we get  $\mu_L=m \omega_{L}^{2} z_{N}^2/2$ where $z_{N}$, the Thomas-Fermi size of the longitudinal trapping potential is given by 
 \begin{equation}
 \label{eq:halfw}
 z_N=\bigg[  \frac{3(N-1)g_{11}\eta_T}{2 m \omega_{L}^{2}} \bigg]^{1/3}
 \end{equation}
 In the regime where the Thomas-Fermi approximation holds we obtain,
\begin{equation}
	\label{eq:etal1}
	\eta_L=\int dz |\phi_{N}|^4 = \frac{2}{5} \bigg[ \frac{9\pi m \omega_{L}^{2} \rho_{0}^{2}}{2 (N-1) g_{11}}\bigg]^{\frac{1}{3}} 
\end{equation}
Note that $g_{11}$ on which $\gamma_{1}$ depends on appears in the above equations for $\eta_{T}$ and $\eta_{L}$. However as mentioned before, we assume that $g_{11}$ is known and even though we are treating $\gamma_{1}$ as the parameter that is being measured we are really estimating $g_{22}$ (or equivalently $a_{22}$). Using equations (\ref{eq:etat1}) and (\ref{eq:etal1}) we can have an estimate of $\eta_{N}$ and use it to compute the expected frequency of oscillation of the atomic populations, $\Omega_{N}$. In Figure~\ref{fig:figure1} the numerically obtained value of $O_{I}(t) = {\rm Im} \langle \psi_{N,1} | \psi_{N,2} \rangle$ is compared with the expected signal $T_{1}(t) = \sin \Omega_{N}t$ for two different values of $N$. We see that the agreement is reasonable for very short times but breaks down quickly.

\begin{figure}[!htb]
\resizebox{8 cm}{5 cm}{\includegraphics{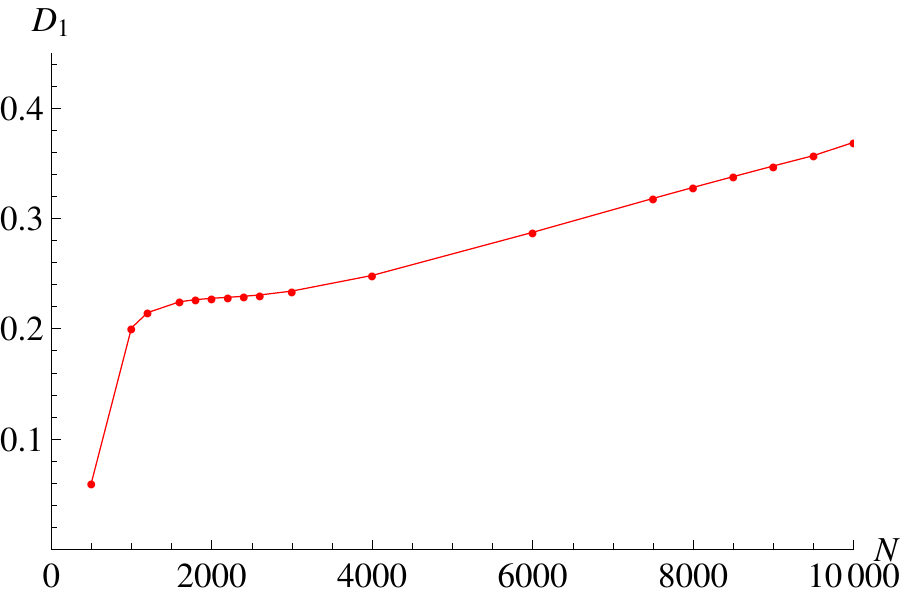}}
\caption{(color online) $D_{1}$ versus $N$ for an anisotropic BEC held in a trap with transverse frequency 350Hz and longitudinal frequency 3.5Hz. $D_{1}$ is slowly increasing with N.}
\label{fig:figd1} 
\end{figure}

The assumption that the transverse part of the wave function is just the ground state wave function of the harmonic trap is valid in the low $N$ limit, while the assumption that the longitudinal wave function is given by the Thomas-Fermi approximation is valid in the large $N$ limit. We therefore expect that the best agreement between the expected signal and the numerical one will be at an intermediate value of $N$ between $N_{L}$ and $N_{T}$ provided the assumption that the wave function has the product form in Eq.~(\ref{eq:gnd1}) does not break down badly in this regime. To compare the agreement between theory and numerics we use as a measure the root-mean-square deviation of the expected signal from the numerically obtained one averaged over a single ``period'' of the expected signal, i.e. 
\begin{equation}
	\label{eq:compare1}
	D_{1}(N) = \sqrt{\frac{1}{m} \sum_{i=1}^{m;\, t_{m} \leq \tau } \Big[ O_{I}(t_{i}) - T_{1}(t_{i}) \Big]^{2}  } ,
\end{equation}
where $\tau$ is the time between two successive zero crossings in the same direction of the expected signal and $t_{i}$ are the time values at which the overlap $O(t)$ has been numerically computed. The measure $D_{1}(N)$ as a function of $N$ is shown in Figure~\ref{fig:figd1}. We see that the deviation is slowly increasing with N. There is no optimal intermediate value of $N$ for which the agreement is best in this case.  

\subsection{Position dependent phase}

\begin{figure}[!htb]
\centering
\resizebox{8 cm}{5 cm}{\includegraphics{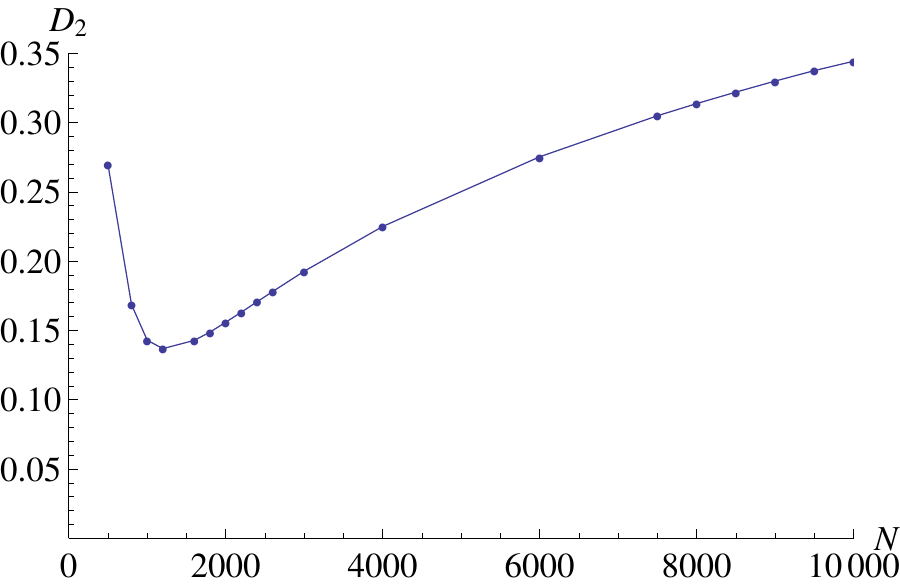}}
\caption{(color online) $D_{2}$ versus $N$ corresponding to an anisotropic BEC in a trap with transverse frequency 350Hz and longitudinal frequency 3.5Hz. We see that the position dependent phase approximation is quite good over a period of the expected signal, for a range of values of $N$.}
\label{fig:figure2} 
\end{figure}

Because of the difference in scattering lengths, $a_{11}$, $a_{22}$ and $a_{12}$, atoms in each of the two internal states see slightly different effective potentials due to scattering even if it is arranged so that both sets of atoms see the same external trapping potentials. The assumption that the spatial profile of the wave function of both sets of atoms is identical breaks down very quickly because of the different potentials seen by the atoms and at long enough times, the two sets of atoms end up segregating~\cite{hall_dynamics_1998}. Modeling the segregation of the atoms within the mean field approximation may not be consistent since the differential velocities acquired by atoms of each type may give at least some of them enough kinetic energy to go out of the ground state. So we will not attempt to track down the effect of the segregation of atoms on the metrology protocol. However, prior to the segregation itself, the wave functions of the two modes picks up position dependent phases. Since the differences in the scattering energy are negligible compared to the transverse trap depth, we assume that the position dependent phase develops only in the longitudinal part of the wave function. Keeping the distribution of atoms identical in the longitudinal direction also, 
\[ |\phi_{N}(z, t)|^{2} = q_{0}(z), \]
we obtain the position dependent relative phase between the two modes as~\cite{boixo_quantum-limited_2009}
\begin{equation}
	\label{eq:rel2}
	\delta \theta(z) = \Omega_{N}t \bigg( 1 + \frac{q_{0}(z) - \eta_{L}}{\eta_{L}} \bigg),
\end{equation}
The overlap integral that gives the expected signal is then given by
\begin{equation} 
	\label{eq:T2}
	T_{2}(t) = {\rm Im} \Big[ e^{-i \Omega_{N}t} \int dz \, q_{0} e^{-i \Omega_{N}t (q_{0} - \eta_{L})/\eta_{L}} \Big]. 
\end{equation}
The $N$ dependence of the deviation $D_{2}(N)$ defined using Eqs.~(\ref{eq:T2}) and (\ref{eq:compare1}) is plotted in Figure~\ref{fig:figure2}.

The assumptions that led to the expression for $T_{2}(t)$ in Eq.~(\ref{eq:T2}) again include a product form,
\begin{equation}
	\label{eq:gnd2} 
	\psi_{N, \alpha} = \chi_{0}(\rho) \phi_{N, \alpha} (z, t), \qquad \alpha = 1,2. 
\end{equation}
Further $\chi_{0}$ is again the ground state wave function of the trap and $\phi_{N, \alpha}$ are the wave functions for each of the two modes along the longitudinal direction with the position dependent relative phase. To compute $T_{2}(t)$ we further assume that $q_{0}(z)$ is the square of the Thomas-Fermi wave function. Our expectation that the assumptions about both $\chi_{0}$ and $q_{0}$ become approximately valid at an intermediate range of $N$ is borne out by Fig.~\ref{fig:figure2} where the agreement between the numerically computed signal and the theoretically expected one is best. 

\begin{figure}[!htb]
\resizebox{8 cm}{5 cm}{\includegraphics{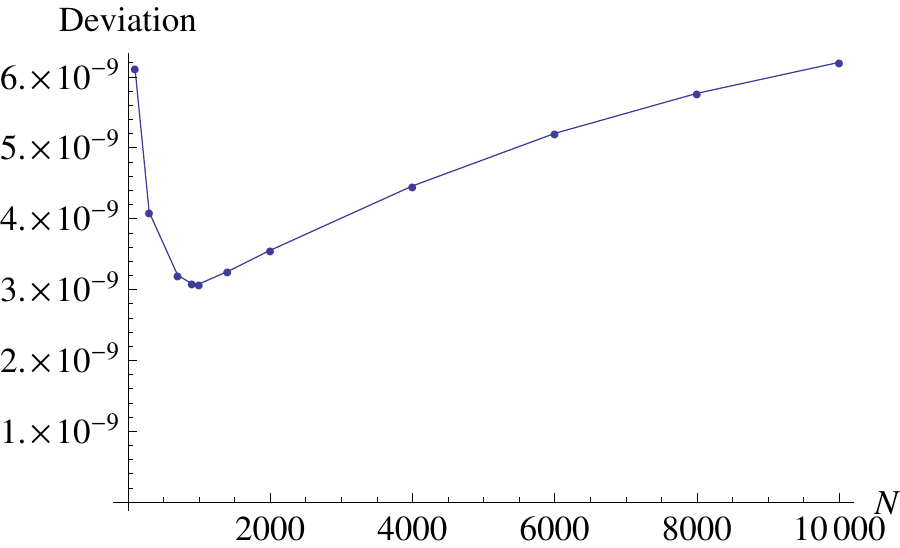}}
\caption{(color online) $N$ dependence  of the RMS deviation of the Thomas-Fermi wave function from the longitudinal (marginal) part of the Gross-Pitaevski ground state. The longitudinal trapping frequency used is 3.5Hz and the transverse frequency is 350Hz.}
\label{fig:figure3} 
\end{figure}

\begin{figure} [!htb]
\resizebox{8 cm}{5 cm}{\includegraphics{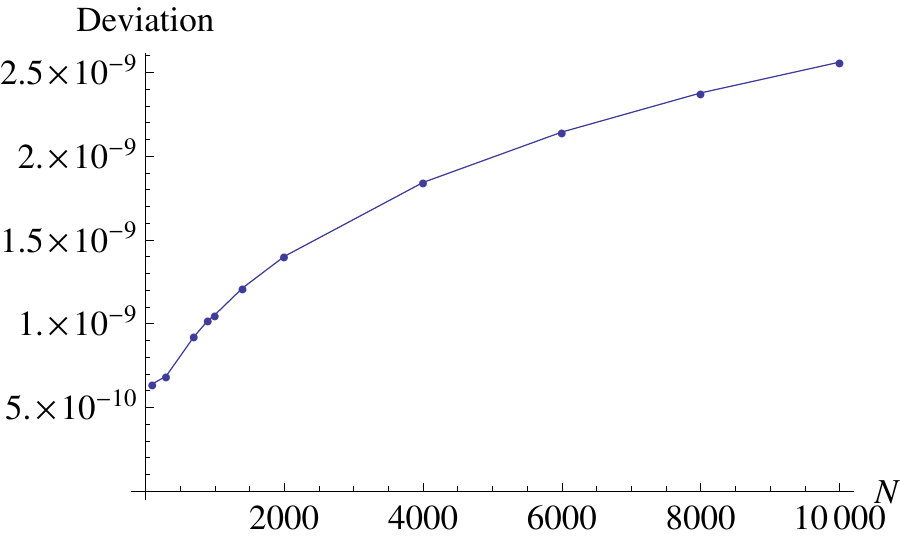}}
\caption{$N$ dependence  of the RMS deviation of the Harmonic oscillator ground state wave function from the transverse (marginal along $x$-axis) part of the Gross-Pitaevski ground state. The longitudinal trapping frequency used is 3.5Hz and the transverse frequency is 350Hz.}
\label{fig:devx} 
\end{figure}

Including the motion of the atoms and attended change in the spatial profile of the wave functions of the two modes is beyond the scope of the mean field approximation we consider. So we are restricted to a time regime in which such motion is negligible as far as the proposed experiment goes. Within this regime, in order to get a better theoretically expected signal, we have to improve the expressions for $q_{0}$, $\eta_{T}$ and $\eta_{L}$. In Fig.~\ref{fig:figure3}, the RMS deviation of the absolute value square of the longitudinal part of the numerically computed  Gross-Pitaevski ground state wave function for atoms in state $|1\rangle$ from the square of the Thomas-Fermi wave function is plotted as a function of $N$. This again shows that the estimate of $q_{0}$ and $\eta_{L}$ that goes into the computation of $T_{2}(t)$ is off the mark for very low values of $N$ as well as for high values of $N$. 
Similary, in Fig.~(\ref{fig:devx}), the deviation of the transverse part of the GP ground state along the $x$-axis from the harmonic oscillator ground state wave function for the same trapping frequency is shown. Here we see that the deviation monotonically increases as a function of $N$. 

\section{Perturbative corrections to the initial wave function \label{Sec4}}

The corrections that need to be applied to the initial wave function of the form assumed in Eq.~(\ref{eq:gnd2}) can be viewed as the emergence of true three dimensional behavior in the quasi-one dimensional wave function we assume that will make the product form invalid. In~\cite{taclacaves} the emergence of three-dimensional behavior in reduced-dimension BECs trapped by highly anisotropic potentials is studied using a perturbative Schmidt decomposition of the condensate wave function between the transverse and longitudinal directions. In this section we see how this level of sophistication to the theory at the mean field level can introduce corrections of the right type that can fix the deviations seen above between the expected and numerically obtained signals. In~\cite{taclacaves} the product form for the ground state wave function is not completely abandoned but rather it is replaced by a sum of products, with each term in the sum being treated as corrections to the previous one as
\begin{equation}
	\label{eq:schmidt1}
	\psi = \sum_{n=0}^{\infty} \epsilon^{n} \chi_{n}(\rho) \phi_{n}(z),
\end{equation}
where $\psi$ is the solution of the time independent GP equation, 
\begin{eqnarray}
	\label{eq:gp3}
	\mu \psi & = &  \bigg( -\frac{\hbar^{2}}{2m} \nabla_{\rho}^{2}  - \epsilon \frac{\hbar^{2}}{2m} \frac{\partial^{2} \;}{\partial z^{2}} + \frac{1}{2} m \omega_{T}^{2} \rho^{2} \nonumber \\
	&& \quad  \qquad+\; \epsilon \frac{1}{2}  m \omega_{L}^{2} z^{2} + \epsilon (N-1)g |\psi|^{2} \bigg) \psi.
\end{eqnarray}
The form for $\psi$ in Eq.~(\ref{eq:schmidt1}) can equivalently be viewed as a Schmidt decomposition~\cite{schmidt,schmidt2} with $\{\chi_{n} \}$ and $\{\phi_{n}\}$ forming the Schmidt basis in the transverse and longitudinal directions respectively. In (\ref{eq:schmidt1}), the Schmidt decomposition has been re-written as an expansion in powers of $\epsilon$ by absorbing the Schmidt coefficients, $c_{n}$, that appear in the decomposition into the transverse wave functions so that they are normalized as $\langle \chi_{n} | \chi_{m} \rangle = c_{n}^{2} \delta_{nm}$. The longitudinal wave functions are delta function normalized. In the perturbation theory developed in~\cite{taclacaves}, the chemical potential as well as the the Schmidt basis functions are expanded in powers of $\epsilon$ as
\begin{eqnarray}
	\mu & = & \sum_{m=0}^{\infty} \epsilon^{m} \mu_{m}, \nonumber \\
	\chi_{n} & = & \sum_{m=0}^{\infty} \epsilon^{m} \chi_{nm}, \nonumber \\
	\phi_{n} & = & \sum_{m=0}^{\infty} \epsilon^{m} \phi_{nm}. 
\end{eqnarray}

If we include corrections to first order to the product wave function in Eq.~(\ref{eq:gnd2}), we have
\begin{eqnarray*}
	\psi_{1}(\rho, z) & = & [\chi_{00} (\rho) + \epsilon \chi_{01} (\rho)][\phi_{00}(z) + \epsilon \phi_{01}(z)] \\
 	&& \qquad \qquad +\;  \epsilon \chi_{10}(\rho) \phi_{10}(z). 
 \end{eqnarray*}
The last term makes the wave function corrected to first order entanglement. We are interested in comparing the expected signal with the position dependent phase in Eq.~(\ref{eq:T2}) with the numerically obtained one when the perturbative corrections are added. When we compute the longitudinal distribution $q_{0}$ by integrating out the transverse part, the contribution from the term containing $\phi_{10}$ is of higher order and so we will not consider this term in the following. Without this term, $\psi_{1}$ retains the product form. 

Consistent with our development so far, we take $\chi_{00}$ to the Gaussian ground state wave function, $\xi_{0}$, of the transverse harmonic trap while $\phi_{00}$ is the longitudinal Thomas-Fermi wave function from Eq.~(\ref{eq:lwave}). We can improve the computation of the expected signal by using a numerical solution to the reduced GP equation in (\ref{eq:redGP1}). However we restrict to the Thomas-Fermi approximation since it allows the theoretical computation to be done without using numerical integration while at the same time allowing us meet the objective of this Paper of seeing how each layer of approximation improves the estimate of the measured parameter, $a_{22}$. The correction to the transverse wave function we consider is given by
\begin{equation}
	\label{eq:trans01}
	\chi_{01} = -\eta_{L}g_{11}(N-1) \sum_{n=1}^{\infty} \xi_{n} \frac{\langle \xi_{n} | \chi_{00}^{3} \rangle}{E_{n} - \mu_{0}},
\end{equation}
where $\xi_{n}$ are the eigenfunctions of the two dimensional, transverse harmonic trap with corresponding energies $E_{n}$ and $\mu_{0} = E_{0}$ where $E_{0}$ is the ground state energy of the transverse trap. Defining
\[ \phi_{0}(z) = \phi_{00}(z) + \phi_{01}(z) \qquad {\rm and} \qquad \tilde{\mu}_{L} = \mu_{L} + \mu_{1}, \]
the first order correction $\phi_{01}$ can be obtained from the reduced Gross Pitaevski like equation for $\phi_{0}$,
\begin{equation}
	\label{eq:long01}
	\bigg[ \frac{1}{2} m \omega_{L}^{2} z^{2} + (N-1)g_{11} \eta_{T} \phi_{0}^{2} - 3 g_{11}^{2} (N-1)^{2} \Gamma_{T} \phi_{0}^{4}\bigg] \phi_{0} = \tilde{\mu}_{L} \phi_{0},
\end{equation}
where 
\begin{equation}
	\label{eq:GammaT}
	\Gamma_{T} = \sum_{n=1}^{\infty} \frac{\langle \xi_{n} | \xi_{0}^{3} \rangle^{2}}{E_{n} - \mu_{0}}. 
\end{equation}
Note that in Eq.~(\ref{eq:long01}) we have ignored the kinetic energy term to be consistent with the Thomas-Fermi wave function we are using for $\phi_{00}$.

\begin{figure}[!htb]
\resizebox{7.5 cm}{4.6 cm}{\includegraphics{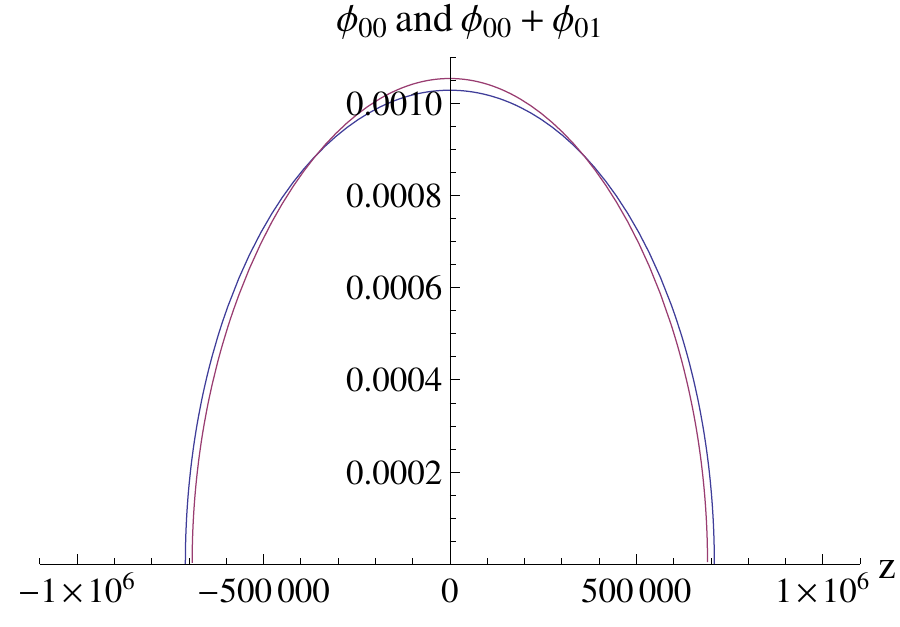}}
\resizebox{7.5 cm}{4.6 cm}{\includegraphics{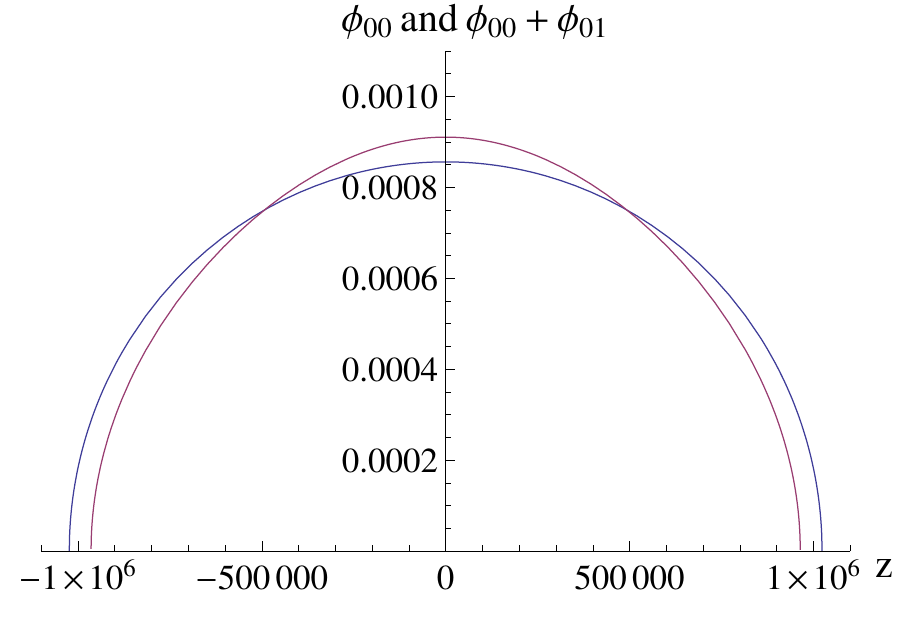}}
\caption{(color online) Corrected (red) and uncorrected (blue) longitudinal wave functions for BECs of $N=1000$ (above) and $N=3000$ (below) atoms respectively held in an anisotropic trap with longitudinal frequency of 3.5Hz and transverse frequency of 350Hz. The $z-$axis is in atomic units.  \label{fig:figure5}}
\end{figure}

For a cigar shaped BEC, we have from \cite{taclacaves}, 
\[ \chi_{01} (\rho) = -a_{11} \eta_{L} (N-1) \sum_{n_{r} = 1}^{\infty} \frac{\xi_{n_{r} 0} (\rho)}{2^{n_{r}} n_{r}}, \]
where $n_{r}$ is the radial quantum number that appears when the eigenfunctions of the two dimensional, transverse, harmonic potential is written in plane polar coordinates. The eigenfunctions with azimuthal quantum number $m$ equal to zero which have finite overlap with ground state wave function and and its powers are given by
\[ \xi_{n_{r}0} = e^{-\rho^{2}/2 \rho_{0}^{2}} L_{n_{r}}(\rho^{2}/\rho_{0}^{2}) \sqrt{\pi} \rho_{0}, \]
where $L_{n_{r}}(x)$ are the Laguerre polynomials. For a cigar shaped trap, we also have, 
\[ \Gamma_{T} =  \frac{\eta_{T}^{2}}{2 \hbar \omega_{T}} \ln \frac{4}{3}. \]
The algebraic, fourth order equation (\ref{eq:long01}) for $\phi_{0}$ has solutions
\[ (\phi_{0})^{2} = \frac{\eta_{T} \pm \eta_{T} \big[ 1 - 12 \Gamma_{T} \eta_{T}^{-2} (\tilde{\mu}_{L} - m\omega_{L}^{2} z^{2}/2 )\big]^{1/2}}{6 g_{11} (N-1) \Gamma_{T}}. \] 
The solution with the minus sign is consistent with the requirement that when the term containing $\Gamma_{T}$ in Eq.~(\ref{eq:trans01}) is absent, the solution reduces to the Thomas-Fermi wave function in Eq.~(\ref{eq:lwave}). The normalization of $\phi_{0}$ is used to find the unknown quantity $\mu_{1}$ that determines $\tilde{\mu}_{L}$. From $\phi_{0}$ we get $\phi_{01} (z) = \phi_{0}(z) - \phi_{00} (z)$.  In Fig.~(\ref{fig:figure5}), plots of $\phi_{0}$ and the correction $\phi_{01}$ are shown and we see that for atom numbers between $N_{L}$ and $N_{T}$, the correction narrows down the Thomas-Fermi wave function as expected and consequently increases $\eta_{L}$. We also see that because we are keeping only the first order correction to the longitudinal wave function, there is a tendency to over-correct the wave function as $N$ increases as noted in~\cite{taclacaves}. This can be mitigated by going to higher orders but then the wave function will not remain separable between the transverse and longitudinal dimensions. 

\begin{figure}[!htb]
\resizebox{8 cm}{5 cm}{\includegraphics{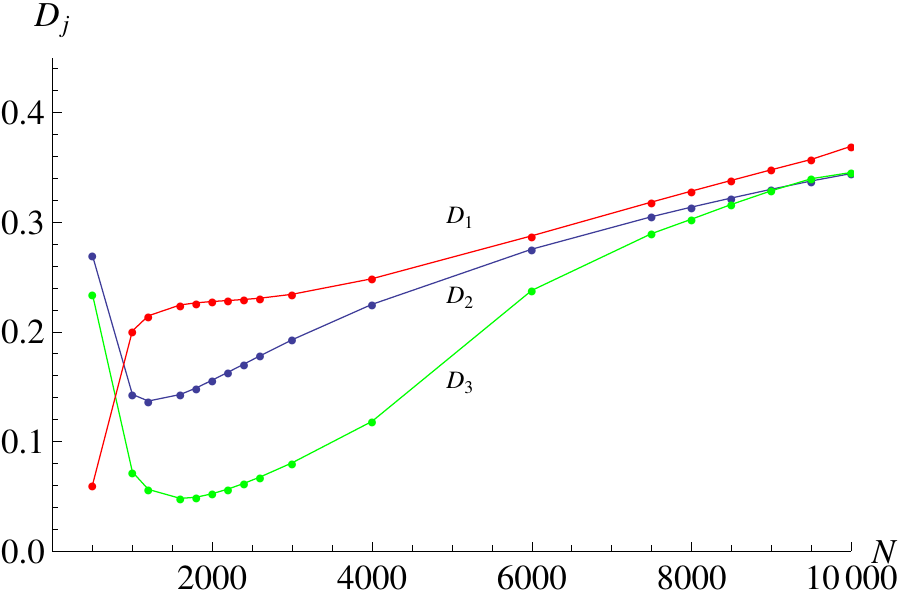}}
\caption{(color online) Comparison of the $N$ dependence of the deviation of the theoretically expected signals $T_{1}$, $T_{2}$ and $T_{3}$ from the numerically simulated data. The red line shows $D_{1}$ as a function of $N$, the blue line is $D_{2}$ and the green line is $D_{3}$. We see that when the corrections due to the position dependent phase as well as the perturbative corrections are added in $T_{2}$ and $T_{3}$ respectively, the deviations progressively decreased. The trap geometry used is 350Hz as the transverse frequency and 3.5Hz as the longitudinal frequency. }
\label{fig:figure6}
\end{figure}

With $\eta_{T}$ and $\eta_{L}$ computed as
\begin{eqnarray*}
	\eta_{T} & = &  \int d \rho \, |\chi_{00} (\rho) + \chi_{01}(\rho) |^{4} , \\
	\eta_{L} & = & \int dz\, |\phi_{00}(z) + \phi_{01} (z) |^{4},
\end{eqnarray*}
 we can compute the expected signal with position dependent phase as
\begin{equation} 
	\label{eq:T3}
	T_{3}(t) = {\rm Im} \Big[ e^{-i \Omega_{N}t} \int dz \, |\phi_{0}|^{2} e^{-i \Omega_{N}t (|\phi_{0}|^{2} - \eta_{L})/\eta_{L}} \Big]. 
\end{equation}
The $N$ dependence of the deviation $D_{3}(N)$ defined using Eqs.~(\ref{eq:T3}) and (\ref{eq:compare1}) is compared with $D_{1}(N)$ and $D_{2}(N)$ in Figure~\ref{fig:figure6}.

We see that adding the perturbative correction further reduces the cumulative RMS deviation between the theoretically expected signal and the actual one. The smaller deviation translates to a better estimate of $a_{22}$. The deviation can be further reduced by dropping the Thomas-Fermi approximation and solving the reduced GP equation for the longitudinal wave function numerically with and without the perturbative corrections. 

\section{Conclusion \label{concl}}

The main question we have addressed in this paper is the choice of function with one free parameter to fit the data from a proof-of-principle quantum metrology experiment using a two mode BEC in a highly anisotropic, cigar shaped trap. We have looked at three analytical models based on different simplifying assumptions that give  theoretically expected signals which when fitted to the observations will yield the value of the measured parameter. We computed the deviation of the theoretically expected signal from simulated data produced by numerically integrating the coupled GP equation describing the system. In the numerical integration of the GP equation we assign a value to the measured parameter. We computed the theoretical signal for the same value of the parameter and quantified the deviation between the theoretical and simulated curves by taking root-mean-squared difference between the two.  We showed that the perturbative approach to finding the initial state of the BEC in the anisotropic trap proposed in~\cite{taclacaves} led to a better theoretical fit for certain ranges of atom numbers in the BEC. It is possible to have a hybrid approach as in~\cite{Boixo-Tacla} and use the values of $\eta_{T}$ and $\eta_{L}$ obtained from the numerically computed initial state of the BEC, which does not depend on $a_{22}$, in Eq.~(\ref{eq:rel2}). However our focus is on how well the theoretical models predict the behavior of the metrology setup and hence we do not include this approach in the present discussion.  In~\cite{taclacaves2} extending the perturbative approach to obtain corrections to the time evolution of the two mode BEC is discussed. The measured parameter appears in the perturbative equations themselves and not just in the solutions which are theoretically expected signals. Going beyond the results presented in this Paper, the theoretically expected signal can be further improved by solving the perturbative equations in~\cite{taclacaves2} treating $a_{22}$ as a fitting parameter.

\acknowledgments
The authors thank Alexandre B. Tacla for a critical reading of the manuscript and valuable comments. This work is supported in part by a grant from the Fast-Track Scheme for Young Scientists (SERC Sl.~No.~2786), and the Ramanujan Fellowship programme (No.~SR/S2/RJN-01/2009), both of the Department of Science and Technology, Government of India. 

\bibliography{bib}

\end{document}